# Female Entrepreneur on Board:

# Assessing the Effect of Gender on Corporate Financial Constraints

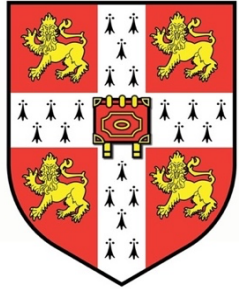

## XIAO Ruiying

### The Faculty of Economics

### The University of Cambridge

Part IIA Paper 3 Theory and Practice of Econometrics I

## Abstract

This study investigates the impact of female leadership on the financial constraints of firms, which are publicly listed entrepreneurial enterprises in China. Utilizing data from 938 companies on the China Growth Enterprise Market (GEM) over a period of 2013-2022, this paper explores how the female presence in CEO positions, senior management, and board membership influences a firm's ability to manage financial constraints. Our analysis employs the Kaplan-Zingales (KZ) Index to measure these constraints, encompassing some key financial factors such as cash flow, dividends, and leverage. The findings reveal that companies with female CEOs or a higher proportion of women in top management are associated with reduced financial constraints. However, the influence of female board members is less clear-cut. Our study also delves into the variances of these effects between high-tech and low-tech industry sectors, emphasizing how internal gender biases in high-tech industries may impede the alleviation of financing constraints on firms. This research contributes to a nuanced understanding of the role of gender dynamics in corporate financial management, especially in the context of China's evolving economic landscape. It underscores the importance of promoting female leadership not only for gender equity but also for enhancing corporate financial resilience.

# 1. Introduction

The dynamic interplay between gender roles and corporate management has emerged as a pivotal area of scholarly focus (Farrell & Hersch, 2005; Francoeur et al., 2008; Wilson Jr, 2014). This burgeoning interest is evident in an array of studies examining the impact of gender on various business behaviors, including corporate governance (Larkin et al., 2012; Boulouta, 2013), entrepreneurship (Schwartz, 1976; Galloway & Sang, 2015), stock performance (Dobbin & Jung, 2010; Gul, et al., 2011) and competitive behaviors (Niederle & Vesterlund, 2007; Betz et al., 1989). Within this broad spectrum, an essential strand of inquiry has been the gender-specific influences on corporate finances, as highlighted by Bøhren & Staubo (2014) and Faccio et al. (2016). Existing literature predominantly discusses inherent traits often ascribed to women in executive roles, such as risk aversion (Maxfield et al., 2010), and a reluctance to exhibit overconfidence (Huang & Kisgen, 2013), as well as the persistent issue of gender discrimination in the financing sector. These traits, often observed in female CEOs and top executives, are linked to more cautious investment and merger strategies, potentially enhancing corporate performance and reducing stock price volatility (Wagana & Nzulwa, 2017; Amore & Garofalo, 2016). Conversely, studies by Muravyev et al. (2009) and Bellucci et al. (2011) point out the ongoing gender bias and stereotype in the financing and banking sector. For instance, female entrepreneurs often face more significant challenges in external investments, encountering financial obstacles related to banking loans (Buttner & Rosen, 1992; Mayoux, 1993), interest rates (Sara, 1998), and venture capitals (Greene et al., 2001; Brush et al., 2018).

However, there is a notable gap in the current body of research in addressing the specific effects of female leadership on corporate financial constraints. Financial constraints, in this context, refer to the barriers that prevent an enterprise from funding all its necessary investments externally (Almeida & Campello, 2007). Such constraints can derive from a variety of sources, such as credit restrictions, inability to secure loans, challenges in issuing equity, or a lack of asset liquidity (Lamont et al., 2001). Further elaboration on this concept will be discussed in the theories and hypothesis section. Yet, it is noticeable that the current literature does not sufficiently address how female leadership specifically influences these financial constraints. For instance, studies by Ruiz-Palomo et al. (2022) and Zhu et al. (2023) have only primarily focused on the moderating effects of gender within the aspects of technological and management innovation, rather than on financial constraints directly.

Henceforth, there are several research gaps in the current body of literature. Firstly, even though past research has examined the female influences in specific corporate roles, such as Board of Directors, CEOs, and top executive positions, there is a dearth of integrated research that carefully examines how these roles collectively influence a company's financial strategies and its ability to manage financial constraints. For example, the work of Zhu Jigao et al. (2012) highlights that companies with a higher ratio of female directors often choose to reduce long-term borrowing during financial crises. This strategic choice helps conserve debt-financing capabilities, setting a foundation for future investment opportunities. Similarly, Pucheta-Martínez et al. (2016) in their Spanish study found that the presence of more women, especially as independent directors on audit committees, significantly enhances the quality of financial reporting. These findings suggest that female directors bring unique perspectives and decision-making approaches, particularly in managing financial risks and reporting. In contrast, research focusing on female executives, like the study by Bellucci et al. (2011), indicates that women in these roles often have less access to social capital and networking opportunities compared to their male counterparts. This disparity affects their ability to secure credit, leading to higher financial constraints due to reduced borrowing capacity. However, these studies, while insightful, do not provide a comprehensive picture. Hence, this research will contribute to the literature by concurrently examining the impact of female leadership across the spectrum—on Boards of Directors, in CEO positions, and within top executive roles—on a company's financing constraints. It would delve into whether these influences are complementary or divergent and how they collectively shape a company's approach to financial management. This kind of holistic analysis is crucial for understanding the multifaceted role of gender in corporate governance and financial strategies, offering valuable insights for both academia and industry practice.

Secondly, the impact of female entrepreneurship on financial constraints has not been sufficiently explored, especially beyond the early stages of venture capital. While existing research has demonstrated that women entrepreneurs face discrimination in initial funding stages, there is a noticeable gap in understanding how these dynamics evolve post-IPO. For instance, studies like those by Greene et al. (2001) and Muravyev et al. (2009) have shown that women-led startups often encounter greater difficulties in securing early-stage capital, potentially due to gender biases and lack of social capital within the investment community. This initial funding challenge can set a precedent for a company's long-term financial health and ability to manage constraints (Lins & Lutz, 2016). However, the literature is sparse on how female entrepreneurs navigate and potentially alleviate or exacerbate financial constraints once their companies go public. Consequently, this research will examine if the strategies employed by women in leadership positions are effective after IPO, and how these strategies impact the company's financial capabilities and market performance. This includes exploring whether female-led public companies demonstrate different patterns in information disclosure, agency cost or investment in innovation compared to their male-led counterparts (Wellalage & Locke, 2013). This research would provide a more comprehensive understanding of the role of gender in corporate financial management, extending beyond the initial venture capital stages to a broader corporate context.

Thirdly, the exploration of gender's leadership influence on corporate finance within different national contexts, particularly in China, remains underdeveloped in the scholarly literature. This gap is significant given the varying impacts observed in different countries. In Pakistan, for instance, research by Amin et al. (2022) has shown that female representation on corporate boards can significantly lower agency costs, thus mitigating principal-agent conflicts. Notably, this effect intensifies in boards with a higher proportion of female directors. On the other hand, Jadiyappa et al. (2019) in India found a contrasting trend, associating an increase in agency costs with the appointment of female CEOs, indicating a potentially negative impact. In China, while there are studies like those by Ain et al. (2021) suggesting that female directors on company boards are instrumental in reducing agency costs linked to conflicts of interest and tend to have a stronger supervisory impact in more economically developed areas of China, these findings are not directly tied to the broader context of financial constraints. However, the connection between these reduced agency costs and the wider spectrum of financial constraints remains unexplored. This link is crucial considering that the level and efficiency of corporate financing are key drivers of economic growth, a factor that gains additional significance in developing countries experiencing economic transition (Jia & Zhang, 2011). These countries often contend with higher transaction costs and limited financing avenues due to their evolving institutional frameworks (Liu et al., 2014). Hence, delving into how gender dynamics influence financial constraints across different national landscapes, with a specific focus on China – the world's largest developing economy – is vital. This paper will provide deeper insights for shaping policies and strategies in China that effectively address gender disparities in corporate governance and finance, particularly in the context of emerging markets and transitional economies.

Accordingly, this study investigates the impact of female leadership on the financial constraints of corporations, focusing on publicly listed entrepreneurial enterprises in China. Utilizing data from 938 companies on the China Growth Enterprise Market (GEM) over a period of 2013-2022, we explore how the presence of women in CEO positions, senior management, and board membership influences a firm's ability to manage financial constraints. In this research, the Kaplan-Zingales (KZ) Index is utilized as an advanced metric to evaluate financial constraints. This index includes crucial financial parameters, such as cash flow, dividends, and leverage ratios, which could offer an in-depth understanding of a company's financial status and operational adaptability under the influence of gender-driven leadership. Notably, the results reveal that firms under the guidance of female CEOs or with a significant presence of women in senior management roles generally face reduced financial limitations. This indicates that having women in top executive positions beneficially affects a company's financial nimbleness and robustness. However, the impact of female representation on corporate boards presents a more complex scenario. The data indicates a less definitive correlation between the proportion of female board members and the alleviation of financial constraints, hinting at a more intricate interplay between board composition, corporate governance practices, and financial outcomes. Furthermore, the study delves into sector-specific differences, particularly contrasting the high-tech and low-tech industries. It reveals that internal gender biases in the high-tech sector could potentially hinder the positive influence of female leadership on easing financial constraints. This insight is particularly relevant given the rapidly evolving nature of the high-tech industry and its significant impact on global economic trends.

Overall, this research not only deepens our understanding of the role of gender dynamics in corporate financial management but also highlights the specific nuances of this relationship within the context of China's unique economic environment. By underscoring the beneficial impact of female leadership on corporate financial health, the study advocates for greater gender diversity at the helm of corporate enterprises, not only as a matter of equity but also as a strategic approach to enhance corporate performance and resilience.

## 2. Theories and Hypothesis

### 2.1 Financial Constraints and Corporate Finance

The concept of financial constraints, which has garnered extensive scholarly attention, primarily focuses on the influences of government interventions (Li et al., 2021), external environmental dynamics (Simerly & Li, 2000), and intrinsic corporate characteristics (Filatotchev & Toms, 2006). Such characteristics encompass a range of elements including company size, ownership nature, equity structures, the effectiveness of internal control systems, the quality of accounting information, and the breadth of corporate social responsibility (Whited & Wu, 2006). The strategic decisions made by executive management, pivotal in shaping the financial behavior of corporations, lie at the heart of these discussions (Chu et al., 2016).

To delve into the internal dynamics of financial constraints and their impact on corporate management, a precise assessment of their severity is crucial. The empirical examination of financial constraints can be traced back to the seminal study by Fazzari et al. (1988), which posited the sensitivity of investment to cash flows as a tangible indicator of financial constraints. Building upon this foundation, subsequent research such as Whited and Wu (2006) proposed the Whited and Wu (WW) Constraint Index, along with diverse ranking criteria predicated on corporate characteristics. In addition to this, researchers have sought to infer financial constraints from an array of corporate disclosures, encompassing statements regarding their financing status or changes in investment plans, specific behaviors like the non-payment of dividends, or characteristics such as being small in scale and having low leverage (Farre-Mensa & Ljungqvist, 2016). As such, the genesis of the KZ index is attributed to the research conducted by Lamont et al. in 2001, who, building upon Kaplan and Zingales' classification, estimated an ordered Logit model. This model correlates the degree of financial constraint with five accounting variables, culminating in the construction of the KZ index. The index mirrors the positive impacts of market-to-book ratios and leverage rates, alongside the negative influences of cash flow, dividends, and cash holdings (Hadlock & Pierce, 2010).

This paper leverages KZ index to access the degree of financial constraints. Within empirical research, the KZ index is frequently employed to determine such constraints status of companies as it is particularly esteemed for its multidimensional assessment, which enables a comprehensive evaluation of a firm's financial status (Hadlock & Pierce, 2010). Despite some critiques regarding the index's original model, which intertwines qualitative and quantitative information in its dependent variable – potentially introducing a fixed 'hard-wired element' element due to the mechanical construction of both dependent and independent variables (Hadlock & Pierce, 2010) – the KZ index remains a robust tool. Among other indications, the adaptability of KZ index for longitudinal studies is particularly valuable, allowing researchers to trace the evolution of financial constraints over prolonged periods. This methodology, as applied in influential studies like Hennessy and Whited (2007), also assumes the consistency of financial constraints prevalence across different time periods and business cycles (Campello et al., 2010). The application of the KZ index has proven to be pivotal in elucidating the profound influence of financial constraints on corporate decision-making. The insights derived from employing the KZ index significantly contribute to the understanding of strategic corporate responses under financial limitations, thereby offering valuable perspectives in the fields of corporate finance and economic behavior analysis.

## 2.2 Upper Echelon Theory and Board Gender

The Upper Echelon Theory, as originally proposed by Hambrick and Mason in 1984, posits that the heterogeneity of executive characteristics leads to varied cognitive schemas and behavioral patterns among managers, thereby influencing corporate behaviors and operational decisions (Abatecola & Cristofaro, 2020). This theory has been further explored and validated in various studies, including those by Abatecola and Cristofaro (2020), highlighting the significance of executive traits such as age, gender, educational background, and other personal attributes on corporate business management activities (e.g. Ting et al., 2015; Wang et al., 2016; Chuang, 2007; Bertrand & Schoar, 2003). According to this theory, the characteristics of executive teams serve as proxies for executive cognition, underscoring their vital role in strategic decision-making. The decision-making behavior of executives is influenced by their background traits, enabling executive teams to impact decisions and actions based on their cognitive capabilities and values (Neely et al., 2020).

Scholar interest, particularly in the influence of female ownership or leadership in senior positions and on corporate boards, has been robust (Adams, Haan, Terjesen, & Ees, 2015). The increasing presence of female directors on corporate boards in many countries can be partly attributed to gender quota regulations adopted in over 17 countries and regions, including Australia, Finland, and Italy. These quotas, which vary from 10% to 40%, are often part of good governance guidelines in most countries and have been gradually increasing over the years (Smith, 2018). Therefore, research in this area has predominantly focused on the impact of female representation on various company-level outcomes. Investigations into the correlation between gender diversity and financial performance have yielded mixed results. While some studies identify positive valuation and market responses (Liu et al., 2014; Reguera-Alvarado et al., 2017), others indicate a negative relationship (Ahern & Dittmar, 2012) or mixed financial outcomes (Carter et al., 2003). These discrepancies can be attributed to variations in research methodologies, performance metrics, time frames, geographical samples, and insufficient control for external factors. The following sections will analyze the dynamics behind those diverse results.

## 2.3 Risk aversion and Overconfidence

A critical aspect influencing these dynamics is the intrinsic personality traits of female CEOs and executives, who are generally perceived as less overconfident and more risk-averse compared to their male counterparts (Chen et al., 2019; Palvia et al., 2015). This pattern is reflected in their approach to investments, where the typically cautious confidence of women leads to more judicious investment decisions, subsequently reducing the frequency of investment and merger actions (Levi et al., 2014). Moreover, female CEOs and top executives are often noted for their commitment to higher ethical standards in their decision-making, along with a cautious and risk-averse style (Vermeir & Van 2008). These characteristics enhance the quality of financial reporting and lead to more rigorous auditing practices. Additionally, female board members are less likely to alter financial earnings and reports, thereby lowering the risks linked to legal disputes and damage to reputation (Francis et al., 2015). Consequently, a firm's dedication to gender diversity may serve as an indicator of the integrity of its corporate governance and the robustness of its disclosure practices (Huang & Hung 2013). The greater conservatism, risk aversion, and ethical vigilance of female managers can lead to diminished involvement in opportunistic management practices, thereby ensuring higher quality in financial reporting and reduced information asymmetry (Wahid, 2019). For example, Zhu et al. (2012) observed that firms with a higher proportion of female executives were quicker to cut down investment levels during financial crises, indicating a reduced need for external capital. Furthermore, Female CEOs' inclination towards risk aversion often leads to a preference for less debt financing, especially avoiding short-term debts with higher financial risks (Beckmann & Menkhoff, 2008). Nevertheless, it is also prevalent that female executives typically have less access to social capital and networking opportunities than their male counterparts (Muravyev, 2009), leading to a decreased likelihood of securing credit and a tendency towards lower debt financing, consequently resulting in increased financing constraints. Consequently, this might initially suggest an increased likelihood of financing constraints for such firms. Yet, this assumption needs to be juxtaposed with the recognized risk-averse nature and conservative decision-making practices of female CEOs and executives. In contrast, the behavioral patterns of male managers, particularly those exhibiting overconfidence, demonstrate a preference for debt financing as a strategic choice for leverage. This tendency has been observed in various contexts, including in the Chinese market, where Brick et al. (2006) found that overconfident managers tend to favor debt over equity financing. Wu (2020) reinforced this observation, noting that managerial irrationality, especially overconfidence, significantly impacts financing decisions in companies listed on the Shanghai and Shenzhen stock exchanges. Therefore, despite the initial limitations in social capital and networking, the presence of female executives and CEOs is likely to bring about more prudent investment decisions and a lower propensity for high-risk debt financing. Given these observations, the following hypothesis can be formulated:

**H1: The increased presence of female management in a firm is associated with reduced financial constraints.**

**H2: The presence of female CEOs is negatively associated with the firm's financial constraints.**

## 2.4 Cost agency theory and Information asymmetry

Agency theory underscores the inherent conflicts of interest and variations in risk preferences between agents and principal (Eisenhardt, 1989). A critical issue in this dynamic is information asymmetry, wherein agents, who typically possess more information than principals, may make decisions that do not fully align with the firm's best interests. Addressing this challenge, corporate governance mechanisms play a critical role in bridging the informational gap between internal stakeholders, external investors, and market participants (Holm & Schøler, 2010). Within this framework, the board of directors is a key internal control mechanism, functioning not only as an oversight body but also as a proponent of shareholder interests, a notion supported by the seminal work of Jensen and Meckling (1976).

The diversity within the board, involving members with varied traits, expertise, and competencies, is posited to enhance the board's effectiveness in managerial control and oversight (Khuong et al., 2022). Such diversity is beneficial in mitigating institutional conflicts, as argued by Guariglia & Yang (2016). Empirical research further highlights the distinctive contributions of female directors. Studies by Amin et al. (2022), Hordofa (2023), and Tang (2022) suggest that women on boards provide deeper insights equipping better interpersonal communication skills, exhibit more rigorous scrutiny, and proactively address issues often overlooked by their male counterparts. This heightened vigilance and diversity of perspectives are thought to bolster the board's independence and overall effectiveness. The reduction in agency costs and the diminishment of information asymmetry, through these means, may lead to more judicious decision-making and efficient resource utilization. This, in turn, can enhance a company's creditworthiness and diminish its dependence on external financing, potentially alleviating financial constraints. Thus, our hypothesis is formulated as follows:

**H3: The presence of female directors on boards is associated with a lower extent of financial constraints compared to boards with male counterparts.**

## 3. Methodology

### 3.1 Data and Sample Selection

Considering the issues of information confidentiality in market competition, unlisted enterprises, especially those in the early stages of development, typically do not disclose their operational data publicly. Consequently, this paper strategically confines its research sample to enterprises listed on the China Growth Enterprise Market (GEM) of the Chinese stock market, in order to investigate the impact of women on entrepreneurial firms. Meanwhile, this study also encompasses data from the Guotai Junan Securities database, which provides insights into the financial conditions, corporate governance, and other characteristics of those enterprises.

The data, refined through the exclusion of incomplete entries, encompasses a diverse array of data from 938 companies, amounting to 6196 observations, listed on the GEM across a period of ten years (2013-2022). This selection is strategically focused on the GEM market to yield a profound understanding of the post-IPO financial landscape for Chinese entrepreneurial firms. Diverging from the conventional frameworks of China's main market board and the SME board, the GEM was inaugurated in October 2009 with a strategic focus on high-tech and start-up companies (Yang, 2018) aligning with the Chinese government's aspiration to establish a trading platform mirroring the functionalities of NASDAQ (Yang, 2018). Consequently, this study's emphasis on entrepreneurial

SMEs within the GEM is justified by its distinct stature in China's financial ecosystem (Hu et al., 2021), resonating with precedents in scholarly inquiries, notably those by Hu et al. (2021) and Wang et al. (2012), which have delved into the evolutionary traits of entrepreneurial SMEs in China, leveraging data from this market.

## 3.2  Data Processing

This paper undertakes the following specific treatments of the collected data. Firstly, it excludes business entities labeled "ST" in the Chinese stock market. This designation is applied by the Securities Regulatory Commission to companies consistently reporting losses, signaling potential risks to investors. These entities generally suffer from a lack of investor trust and face significant challenges in securing market financing. Consequently, they are treated as atypical samples and are omitted from the study. Subsequently, the study implements a winsorization technique at the 1% level for all continuous variables. This approach is designed to minimize the impact of extreme data points on the results of the regression analysis.

## 3.3  Variables

### 3.3.1  Dependent Variable: financial constraints

In this research, the Kaplan-Zingales (KZ) Index is utilized to assess the financial limitations faced by listed entrepreneurial firms (Kaplan & Zingales, 1997). The computation of this index involves the extraction of five key variables to gauge a company's financial constraints. These variables encompass: the operational net cash flow to total assets ratio from the prior period, the cash dividends to total assets ratio from the prior period, the cash holdings to total assets ratio from the prior period, the debt-to-asset ratio, and Tobin's Q. The entire sample is ranked based on these five factors, with values assigned relative to the median. For example, if a company's operational net cash flow to the previous period's total assets is below the median, a new variable, KZ1, is generated and assigned a value of 1. This procedure is replicated for each variable, cumulatively adding them to calculate the initial value of the KZ Index. This KZ Index, generated using this methodology, serves as the dependent variable. The aforementioned five variables are used as explanatory variables, and an ordered logit regression model is applied to re-estimate and fit the KZ Index. The newly generated fitted values thus serve as the final dependent variables (kz) in this study to evaluate the level of financial constraints faced by the enterprises. Accordingly, the KZ index is a well-established proxy for gauging a firm's limitations in accessing external finance, reflecting its investment-cash flow sensitivity and other critical financial situations (Almeida & Campello, 2007).

### 3.3.2  Main Independent Variables: female involvement

To quantitatively measure the involvement of women in various managerial processes of enterprises, this study constructs variables based on the roles of CEO, senior management personnel, and board members. These variables include the presence of a female CEO (fceo), the proportion of women in senior management (fmanage), and the proportion of women on the board of directors (fboard), serving as explanatory variables. This variable configuration follows studies such as Gupta et al. (2018), Liu et al. (2014), and Valentine & Rittenburg (2006). However, diverging from the past research, which predominantly focused on the performance of women in singular key positions within enterprises, this study innovatively and comprehensively investigates the permeation of female participation across various managerial processes. Additionally, it explores the varying preferences of different companies in the appointment of female managers. This approach not only highlights the roles of women in high-level positions but also provides a holistic view of their integration into the broader spectrum of corporate management.

### 3.3.3 Control variables

In selecting control variables, this study adheres to the criterion of ensuring a high correlation with the dependent variable while maintaining minimal association with the explanatory variables. To guarantee the systematic and comprehensive selection of control variables, the study embarks on choices from three dimensions: business operation characteristics, financial performance, and governance level. The specific selections are as follows:

Firstly, in terms of operational Characteristics, the study employs 'enterprise size' (size) and 'establishment years' (age) to represent business operation characteristics. The former is calculated using the natural logarithm of total assets of the enterprise, while the latter is determined through the natural logarithm of the years since establishment. Hadlock & Pierce (2010) found a quadratic relationship between enterprise size and its level of financing constraints, and a linear relationship with the number of years since establishment. This theoretical derivation is primarily rooted in the scale effects and life cycle theory of enterprise development. Enterprises in growth and maturity phases typically possess more abundant and stable asset liquidation opportunities, thereby facilitating external financing, particularly during periods of high investor sentiment (Lu & Wang, 2018). On the other hand, enterprises that have achieved scale economies, compared to those that have not, exhibit higher asset liquidation efficiency and are perceived as more valuable investment opportunities from an investor's perspective (Abu & Kirsten, 2009).

Secondly, regarding financial performance, the study incorporates 'return on assets' (ROA) and 'leverage' as indicators. ROA, defined as the net income divided by total assets, reflects the efficiency of asset utilization in generating profits. This metric is crucial in assessing an enterprise's internal operational efficiency and overall financial health (Jordão & Almeida, 2017). A higher ROA often indicates more efficient management and better financial performance, which can affect investment decisions and investor confidence (Omondi & Muturi, 2013). Conversely, the debt ratio, calculated as total debt divided by total assets, represents the financial leverage of a company. The leverage ratio is essential in understanding the risk profile of an enterprise, as higher debt ratio can lead to increased financial distress but may also signal growth opportunities through external financing (Myers, 2001). These financial performance measures are selected based on their strong predictive ability for future growth and stability, as well as their potential impact on external investor behavior (Ittner et al., 2003).

Lastly, in the realm of governance level, the study considers 'board composition', namely the proportion of independent directors on the board, which is a common measure of corporate governance quality. Empirical studies, such as those by Wang et al. (2015), demonstrate that a higher proportion of independent directors correlates with more effective monitoring and improved decision-making, leading to enhanced corporate performance and investor confidence.

## 3.4 Empirical model

To quantitatively assess the impact of gender on corporate financial constraints within entrepreneurial companies, we utilize the fixed effect regression model outlined below:

$$kz_{i,t} = \alpha_0 + \alpha_1 fceo_{i,t} + \alpha_2 size_{i,t} + \alpha_3 roa_{i,t} + \alpha_4 board_{i,t} + \alpha_5 indep_{i,t} + \alpha_6 age_{i,t} + \sum Year + \sum Industry + \varepsilon_{i,t}$$
(1)

$$kz_{i,t} = \alpha_0 + \alpha_1 fmanage_{i,t} + \alpha_2 size_{i,t} + \alpha_3 roa_{i,t} + \alpha_4 board_{i,t} + \alpha_5 indep_{i,t} + \alpha_6 age_{i,t} + \sum Year + \sum Industry + \varepsilon_{i,t}$$
(2)

$$kz_{i,t} = \alpha_0 + \alpha_1 fboard_{i,t} + \alpha_2 size_{i,t} + \alpha_3 roa_{i,t} + \alpha_4 board_{i,t} + \alpha_5 indep_{i,t} + \alpha_6 age_{i,t} + \sum Year + \sum Industry + \varepsilon_{i,t}$$
(3)

In this analysis, the subscripts i and t represent firm i in the year t, respectively. The dependent variable, $kz_{i,t}$, signifies the financial constraints of a firm. The primary independent variable under consideration is 'female representation,' which is quantified using three metrics: (1) the count of female CEOs, (2) the ratio of female executives in top management, and (3) the ratio of female board members to the total count of management and board members. Consistent with previous research, this study incorporates various control variables: the size of the firm, Return on Assets (ROA), Leverage, Establishment Year, and Firm Age. Additionally, the regression model accounts for both firm-specific and year-specific fixed effects, and standard errors are clustered at the firm level for more precise estimation. These fixed effects are incorporated to capture the unobservable heterogeneities in financial constraints that are specific to the enterprise's year of operation and industry sector. For instance, the 2020 U.S. stock market experienced a circuit breaker crisis that generally narrowed financing channels for listed enterprises during that year. Comparatively, enterprises in the high-technology sector are perceived to have more favorable market prospects than those in low-technology sectors, thereby typically accessing a wider range of financing opportunities. In the regression outcomes, this study principally examines the significance and sign of the estimated coefficients, particularly $\alpha_1$, of the core explanatory variables to assess whether the participation of women in corporate governance effectively mitigates the financial constraints encountered by enterprises.

To ensure the feasibility of applying this model, the study conducts a Hausman test to verify the appropriateness of including fixed effects. The Hausman test statistics for the three fixed-effects models, using the variables of whether the CEO is female (fceo), the proportion of women in senior management (fmanage), and the proportion of women on the board (fboard) as explanatory variables, are 584.08, 585.54, and 582.42, respectively, with a significance level of 1%. These results suggest that fixed effects offer a superior model estimation method over random effects. Furthermore, in the empirical results section, the study will further validate the model design's rationality through a robustness check and by observing the goodness of fit.

## 4. Empirical Result

### 4.1 Descriptive Statistics

The descriptive statistical results of the variable data, meticulously collated as per the preceding text, are presented in **Table 1**. This table initially reveals a significant volatility in the level of financing constraints faced by currently listed entrepreneurial enterprises in China. The mean value of the dependent variable, the extent of financing constraints (kz), is 0.595, with a standard deviation markedly exceeding the mean at 2.466. This phenomenon suggests, firstly, that public stock offerings through securities exchanges do not wholly ameliorate the financing challenges of entrepreneurial firms, as a substantial number of such listed enterprises continue to grapple with financing distress. Secondly, there is a noticeable disparity in the ability of these enterprises to handle financing constraints, indicative of an embryonic investment environment in the GEM that has yet to establish a stable channel of capital provision for entrepreneurial firms. Furthermore, upon deeper examination of the data characteristics of the explanatory variables, it is observed that in Chinese listed enterprises in the GEM, the average probability of a female serving as CEO is approximately 5%, with women constituting only 20.2% of senior management and 17.3% of board members. Additionally, the variables representing the proportion of female executives (fmanage) and female board members (fboard) do not exhibit significant volatility in relation to the dependent variable of financing constraints (kz). These results point to a prevalent gender bias within Chinese entrepreneurial enterprises, where, particularly in listed entrepreneurial firms, the probability of women being appointed in senior management roles is comparatively low, and women face greater challenges in gaining recognition for their capabilities and experience (Bishop et al., 2005; Kuhn & Shen, 2012). In addition, Table 1 shows that the enterprises in our sample have an average firm size of 21.54, ROA of 2.9%, board size of 2.047, number of independent boards of 38.5%, and firm age of 2.851. In summary, the distribution characteristics of the selected dependent and explanatory variables suggest that currently listed entrepreneurial enterprises in China have considerable room for maneuver in addressing financing constraints. Concurrently, there is a substantial need to rectify the biases against women's competencies within these enterprises. Hence, researching the strategic channels through which Chinese listed entrepreneurial firms can alleviate financing distress, as well as the participation of women in corporate management, holds significant scholarly relevance and practical necessity.

**Table 1. Descriptive Statistics.**

| Variable | Obs | Mean | Std. Dev. | Min | Max |
| --- | --- | --- | --- | --- | --- |
| kz | 6196 | 0.595 | 2.466 | -6.324 | 6.136 |
| fceo | 6196 | 0.051 | 0.219 | 0.000 | 1.000 |
| fmanage | 6196 | 0.202 | 0.179 | 0.000 | 0.667 |
| fboard | 6196 | 0.173 | 0.140 | 0.000 | 0.571 |
| size | 6196 | 21.54 | 0.838 | 19.959 | 24.003 |
| roa | 6196 | 0.029 | 0.088 | -0.402 | 0.198 |
| board | 6196 | 2.047 | 0.189 | 1.609 | 2.398 |
| indep | 6196 | 0.385 | 0.054 | 0.333 | 0.571 |
| age | 6196 | 2.851 | 0.284 | 2.079 | 3.434 |

Note: Table 1 shows presents a summary of key data for the main variables in our study. The dataset includes 6,196 observations of firm data from publicly traded Chinese companies spanning 2013 to 2022 on the China Growth Enterprise Market (GEM). This study has applied a winsorization process at the 1% level to all continuous variables.

## 4.2 Correlation Matrix

In this research, the Pearson correlation test is utilized to avert the potential issue of multicollinearity within the specified model framework. The findings, delineated in **Table 2**, demonstrate that aside from the dependent variable (kz), the absolute values of the correlation coefficients among the other variables do not exceed 0.7, even when statistically significant. This suggests a lack of strong correlations between the explanatory and control variables, mitigating the risk of multicollinearity adversely impacting our model. The findings also imply a significant degree of independence among the control variables, which allows for the individual effects of each variable to be interpreted without the confounding influence of other variables in the model. The selection of control variables is thus validated, underscoring its efficacy in facilitating a robust understanding of the factors influencing the dependent variable, without succumbing to the pitfalls of redundancy or overfitting.

On the other hand, the correlation coefficient between the variable indicating whether the CEO is female (fceo) and the financing constraint index (kz) is -0.035, significant at the 1% level. This result serves as preliminary validation of the hypothesis proposed in this study. Additionally, the correlation coefficients between the other two explanatory variables, the proportion of women in senior management (fmanage) and on the board of directors (fboard), and the dependent variable of corporate financing constraints (kz) are not significant. The results of the correlation test cannot be used to infer causality between these factors, but they do indicate that there is no direct correlation between the presence of women in senior management and board positions and the level of financing constraints faced by firms. This study posits that this phenomenon may be attributed to potential selection bias in the research design. Specifically, an increase in financing constraints might be one of the internal motivations for firms to incorporate women into their management and board (McGuinness et al., 2015; Orazalin, 2019). In other words, listed entrepreneurial firms may proactively optimize the gender structure of their management to address their financing constraint challenges. To ensure the robustness of the results of this study, this issue will be further discussed in detail in the section on robustness checks.

**Table 2. Descriptive Statistics.**

| Variables | (1) | (2) | (3) | (4) | (5) | (6) | (7) | (8) | (9) |
|---|---|---|---|---|---|---|---|---|---|
| (1)kz | 1.000 | | | | | | | | |
| (2)fceo | -0.035 | 1.000 | | | | | | | |
| | (0.007) | | | | | | | | |
| (3)fmanage | -0.017 | 0.095 | 1.000 | | | | | | |
| | (0.186) | (0.000) | | | | | | | |
| (4)fboard | 0.001 | 0.196 | 0.333 | 1.000 | | | | | |
| | (0.906) | (0.000) | (0.000) | | | | | | |
| (5)size | 0.129 | -0.044 | -0.046 | -0.018 | 1.000 | | | | |
| | (0.000) | (0.001) | (0.000) | (0.163) | | | | | |
| (6)roa | -0.509 | 0.004 | -0.008 | -0.002 | 0.016 | 1.000 | | | |
| | (0.000) | (0.745) | (0.529) | (0.904) | (0.197) | | | | |
| (7)board | -0.008 | -0.020 | -0.070 | -0.049 | 0.131 | 0.050 | 1.000 | | |
| | (0.552) | (0.115) | (0.000) | (0.000) | (0.000) | (0.000) | | | |
| (8)indep | 0.033 | 0.051 | 0.056 | 0.038 | -0.056 | -0.040 | -0.658 | 1.000 | |
| | (0.009) | (0.000) | (0.000) | (0.003) | (0.000) | (0.002) | (0.000) | | |
| (9)age | 0.086 | 0.037 | 0.073 | 0.082 | 0.094 | -0.065 | 0.033 | -0.011 | 1.000 |
| | (0.000) | (0.003) | (0.000) | (0.000) | (0.000) | (0.000) | (0.009) | (0.389) | |

Note: Table 2 provides a detailed account of the inter-variable relationships within our dataset, employing the Pearson correlation coefficient as the measure of choice. Our dataset encompasses 6,196 firm observations from publicly traded companies on the China Growth Enterprise Market (GEM), covering a period from 2013 to 2022. The matrix includes both the correlation coefficients and the corresponding p-values (in parentheses) to assess the statistical significance of the correlations.

## 4.3 Financial Constraints and Gender

In this section, the main results present in Table 3. The models' goodness-of-fit indices surpass the 30% threshold, thereby signifying that these fixed-effects models cumulatively elucidate over 30% of the variation in the dependent variable, namely the extent of financing constraints encountered by the firms under study. This substantial explanatory capacity imbues the regression results pertaining to each explanatory variable within the model with a high degree of credibility and validity. The CEO gender (fceo) and the proportion of women in top management (fmanage), the coefficients are -0.318 and -0.326, respectively, with significance levels at 1% and 5%. These findings robustly corroborate the research hypotheses 1 and 2 posited earlier, establishing a marked positive causal nexus between the participation of women in executive roles of publicly listed entrepreneurial firms in China and factors such as enhanced accessibility to financing avenues and the magnitude of financial leveraging. In essence, the data indicates that in these enterprises, female leadership in CEO or senior management roles is conducive to mitigating financing constraints and augmenting the dynamics of cash flow outflows. In the context of their financing activities, women frequently demonstrate superior problem-solving and managerial competencies in comparison to their male counterparts.

On the other hand, the estimated coefficient for the proportion of women on the board (fboard) is -0.202, which, while also negative, is not statistically significant. This study speculates that this result may be related to the functional setup of board institutions in China. According to the "The Company Law of the People's Republic of China,"[1] the board's predominant functions encompass making decisions on critical corporate matters via meeting votes—such as sanctioning annual operational strategies and foreign investment ventures—and overseeing the everyday management of executive staff. This law also mandates the inclusion of independent directors, who are not engaged in daily managerial tasks. Given this board structure, it is inferred that, with the exception of the CEO, board members generally do not partake directly in the company's internal management processes (Firth et al., 2007; Liu et al., 2015). Against this backdrop, the influence of board members on corporate financing activities is limited to regular supervision and annual reviews. Therefore, this result leads to the rejection of the original hypothesis 3, suggesting that the involvement of women in the board structures of publicly listed entrepreneurial firms in China does not significantly impact the alleviation of current financing constraints for these enterprises.

---

[1] The Company Law of the People's Republic of China, enacted by the National People's Congress of the Chinese government on December 29, 1993, became effective from July 1, 1994. It has undergone several amendments over the years, with the latest version being implemented in 2018. This law governs the operations of limited liability and joint stock companies in China.

Synthesizing the outcomes derived from the three explanatory variables, it is concluded that publicly listed entrepreneurial firms in China can elevate managerial efficacy and alleviate financing burdens by fostering greater female involvement in their management echelons. Nonetheless, this strategic direction hinges on the direct engagement of women in the daily operational activities of these firms.

## 4.4 Industry Patterns and Gender

To further analysis, this study primarily conducts a grouped regression by differentiating between high-tech and low-tech publicly listed entrepreneurial companies. This section serves to illuminate the influence of female managerial participation on the financing constraints encountered by firms within these distinct technological echelons. The rationale for this categorization is twofold. Firstly, gender bias against female staff members is presumed to be more pronounced in high-tech companies. This bias, underpinned by entrenched societal stereotypes in China that undervalue women's proficiency in mathematics and scientific disciplines (Schmader, 2002), manifests in a diminished perception of female managers' capabilities in decision-making and compliance within high-tech contexts. Such perceptions inevitably influence the operational dynamics and gender inclusivity in these firms. Second, the exigencies of financing in high-tech sectors are markedly more pronounced due to their intensive investment in research and development. The level of R&D investment is a critical determinant of their technological advancement, and financing

Pembroke College 26 December 2023

constraints pose a significant impediment to the expansion of R&D endeavors and the initiation of novel research projects (Brown et al., 2012; Sena, 2006). This backdrop renders the exploration of the varied impacts of female managerial participation in high-tech versus low-tech sectors on financing constraints both theoretically intriguing and pragmatically pertinent. Empirically, this study employs the 'Strategic Emerging Industries Classification Catalog' in China to demarcate the industries constituting high-tech entrepreneurial firms. The results of the high-tech and low-tech industry group regression, as shown in [Table 4](#), are as follows: in the high-tech group (1-3 columns), the core explanatory variables—whether the CEO is female (fceo), the proportion of women in senior management (fmanage), and the proportion of women on the board (fboard)—have estimated coefficients of -0.166, -0.190, and -0.060, respectively, none of which are significant. In the low-tech group (4-6columns), the estimated coefficients for these variables are -0.788, -0.958, and -0.942, respectively, with significance levels of 1%, 1%, and 10%. These results confirm the theoretical conjecture that gender discrimination against management staff is stronger in high-tech groups, with female managers having lower levels of participation in internal management and a relatively weaker role in alleviating financing constraints faced by the high-tech industries. Therefore, a critical step for high-tech industries is to recalibrate their corporate culture, prioritizing the reduction of internal gender biases against female staff as a cornerstone of their cultural development and operational strategy.

**Table 3. Benchmark Regression.**

| VARIABLES | (1) kz | (2) kz | (3) kz |
|---|---|---|---|
| fceo | -0.318*** | | |
|  | (-2.68) | | |
| fmanage | | -0.326** | |
|  | | (-2.16) | |
| fboard | | | -0.202 |
|  | | | (-1.06) |
| size | 0.270*** | 0.268*** | 0.273*** |
|  | (7.79) | (7.69) | (7.89) |
| roa | -13.563*** | -13.553*** | -13.563*** |
|  | (-34.45) | (-34.48) | (-34.48) |
| board | 0.355* | 0.330* | 0.339* |
|  | (1.95) | (1.81) | (1.86) |
| indep | 1.728*** | 1.677*** | 1.656*** |
|  | (2.76) | (2.68) | (2.65) |
| age | -0.041 | -0.035 | -0.042 |
|  | (-0.36) | (-0.31) | (-0.37) |
| Constant | -6.077*** | -5.935*** | -6.065*** |
|  | (-6.56) | (-6.37) | (-6.54) |
| Observations | 6,194 | 6,194 | 6,194 |
| R-squared | 0.333 | 0.332 | 0.332 |
| Industry FE | YES | YES | YES |
| Year FE | YES | YES | YES |

Robust t-statistics in parentheses
*** p<0.01, ** p<0.05, * p<0.1

Note: Table 3 presents the results of a benchmark regression analysis exploring the relationships between various firm characteristics and financial constraints. The analysis incorporates data from 6,194 firms listed on the China Growth Enterprise Market (GEM), collected from 2013 to 2022. The t-statistics in parentheses provide insight into the statistical significance of these relationships, with significance levels denoted by asterisks. The positive and negative signs of the coefficients suggest varying degrees of influence, with fceo and fmanage showing a strong negative relationship with KZ index. The model controls for industry and year fixed effects (Industry FE and Year FE), ensuring that the results are not confounded by industry-specific or time-specific factors.

**Table 4. Heterogeneity Test.**

| VARIABLES | (1) kz | (2) kz | (3) kz | (4) kz | (5) kz | (6) kz |
|---|---|---|---|---|---|---|
| fceo | -0.166 | | | -0.788*** | | |
|  | (-1.26) | | | (-3.06) | | |
| fmanage | | -0.190 | | | -0.958*** | |
|  | | (-1.13) | | | (-2.87) | |
| fboard | | | -0.060 | | | -0.942** |
|  | | | (-0.28) | | | (-2.11) |
| size | 0.294*** | 0.293*** | 0.296*** | 0.129 | 0.117 | 0.126 |
|  | (7.74) | (7.66) | (7.79) | (1.53) | (1.37) | (1.50) |
| roa | -14.247*** | -14.250*** | -14.253*** | -11.140*** | -10.945*** | -10.989*** |
|  | (-30.25) | (-30.29) | (-30.29) | (-16.49) | (-16.25) | (-16.29) |
| board | 0.385** | 0.373* | 0.383** | 0.209 | 0.089 | 0.047 |
|  | (2.01) | (1.95) | (1.99) | (0.37) | (0.16) | (0.09) |
| indep | 1.821*** | 1.806*** | 1.802*** | 1.399 | 1.095 | 0.886 |
|  | (2.69) | (2.67) | (2.66) | (0.83) | (0.67) | (0.54) |
| age | -0.077 | -0.073 | -0.080 | 0.212 | 0.244 | 0.287 |
|  | (-0.63) | (-0.59) | (-0.64) | (0.78) | (0.89) | (1.04) |
| Constant | -6.625*** | -6.552*** | -6.647*** | -3.157 | -2.433 | -2.649 |
|  | (-6.60) | (-6.49) | (-6.61) | (-1.29) | (-1.00) | (-1.09) |
| | | | | | | |
| Observations | 5,021 | 5,021 | 5,021 | 1,173 | 1,173 | 1,173 |
| R-squared | 0.309 | 0.309 | 0.309 | 0.426 | 0.426 | 0.424 |
| Industry FE | YES | YES | YES | YES | YES | YES |
| Year FE | YES | YES | YES | YES | YES | YES |

Robust t-statistics in parentheses
*** p<0.01, ** p<0.05, * p<0.1

Note: Table 4 presents the findings from a heterogeneity test that examines the differential impacts of firm characteristics on financial constraints across various industries. This analysis utilizes a sample of 6,194 firms, with a focus on distinguishing the effects within high-tech industry (columns 1-3) from other industries (columns 4-6), using the same dataset from the China Growth Enterprise Market (GEM) spanning 2013 to 2022. The model accounts for both industry and year variations through fixed effects (Industry FE and Year FE), which helps isolate the effects of the firm characteristics from those industry and temporal trends.

## 4.5 Robustness Check

### 4.5.1 Propensity Score Matching (PSM)

This section critically addresses the conjecture of selection bias within the research framework, a concern previously articulated in the correlation assessment section. This bias potentially arises from firms, already constrained in financial capabilities, strategically modifying their managerial gender composition to mitigate obstructions in their existing financial channels. Concurrently, variances in corporate culture across different entities contribute to a heterogeneity in the propensity towards female staff appointments (Lewellyn & Muller-Kahle, 2019). Notably, female Chief Executive Officers (CEOs) exhibit a propensity towards elevating female personnel within their organizations, thereby cultivating an environment emblematic of gender parity in both corporate ethos and operational management (Cook & Glass, 2015). In response to these considerations, this study adopts the Propensity Score Matching (PSM) approach, an advanced methodological tool, to rectify potential discrepancies in the initial research design, thereby ensuring the empirical integrity and robustness of the baseline regression outcomes. The PSM is operationalized via kernel density matching, utilizing the gender of the incumbent CEO as a pivotal variable for matching, and incorporating the control variables from the preliminary baseline regression as the matching covariates. The results of the regression analysis, post-adjustment for disparities between firms governed by male and female CEOs using the PSM methodology, are systematically delineated in Table 5. The estimated coefficients for the variables indicating whether the CEO is female (fceo) and the proportion of females in executive positions (fmanage) are -0.325 and -0.324 respectively, with significance levels of 1% and 4% respectively. These results demonstrate that the significance and direction of the explanatory variables are consistent with the baseline regression. This indicates that, even after accounting for potential selection bias in the empirical design, specifically the corporate preference for employing female staff, the conclusion that women's participation in corporate management can enhance a firm's capacity to cope with financial constraints remains valid. In other words, the regression results from the previous text exhibit strong robustness.

**Table 5. Propensity Score Matching (PSM)**

| VARIABLES | (1) kz | (2) kz | (3) kz |
|---|---|---|---|
| fceo | -0.325*** | | |
|  | (-2.74) | | |
| fmanage | | -0.324** | |
|  | | (-2.14) | |
| fboard | | | -0.209 |
|  | | | (-1.09) |
| size | 0.271*** | 0.270*** | 0.275*** |
|  | (7.79) | (7.72) | (7.91) |
| roa | -13.524*** | -13.515*** | -13.524*** |
|  | (-34.20) | (-34.24) | (-34.24) |
| board | 0.386** | 0.354* | 0.365** |
|  | (2.07) | (1.91) | (1.96) |
| indep | 1.988*** | 1.900*** | 1.890*** |
|  | (3.05) | (2.92) | (2.90) |
| age | -0.031 | -0.026 | -0.033 |
|  | (-0.28) | (-0.23) | (-0.29) |
| Constant | -6.300*** | -6.144*** | -6.274*** |
|  | (-6.76) | (-6.55) | (-6.72) |
| Observations | 6,159 | 6,159 | 6,159 |
| R-squared | 0.332 | 0.332 | 0.332 |
| Industry FE | YES | YES | YES |
| Year FE | YES | YES | YES |

Robust t-statistics in parentheses
*** p<0.01, ** p<0.05, * p<0.1

Note: Table 5 performs a robustness check to rectify selection bias concerns, a critical element previously identified in the correlation analysis. To mitigate these biases, the Propensity Score Matching (PSM) method is employed, using the gender of the CEO as a key matching criterion. The findings presented in Table 5, post-PSM adjustment, exhibit coefficients for female CEO presence (fceo) and female management ratio (fmanage) at -0.325 and -0.324 with respective significance levels, aligning with baseline regressions. These coefficients suggest that the participation of women in management roles likely supports a firm's financial resilience, upholding the robustness of the original results against selection bias.

## 4.5.2 Lagged Dependent Variables

In section 5.5.2, the study adopts a lagged explanatory variable to further check the robustness of the research design outcomes. Three primary considerations motivate the use of this methodology. Firstly, the information about management and board members disclosed in the annual reports of Chinese listed companies pertains to the personnel in office at the year's end. In reality, some of these individuals appointed towards the end of the year may exert limited influence on the firm's management level for that year. By lagging the explanatory variables by one year, the study aims to circumvent any biases arising from this disclosure practice. Secondly, this approach is intended to mitigate any potential endogeneity issues within the empirical design. Thirdly, as discussed in the benchmark regression section, the influence of a listed company's board members on its financing activities is primarily executed through regular supervision and annual summaries. Therefore, the board's involvement in these activities can, in fact, impact the subsequent year's financing activities. Hence, lagging the explanatory variables by one period also serves to verify the rationale behind the non-significant estimated coefficients for the proportion of women on the board (fboard) in the benchmark regression results, supplementing the argument that the appointment of women to the board has a lagged effect on corporate financing activities. The regression results of the lagged variables are displayed in Table 6: the estimated coefficients for whether the CEO is female (fceo), the proportion of women in senior management (fmanage), and the proportion of women on the board (fboard) are -0.320, -0.452, and -0.579, respectively, with significance levels of 5%, 1%, and 1%. The coefficient estimates for whether the CEO is female (fceo) and the proportion of women in senior management (fmanage) remain consistent with the benchmark regression, affirming the robustness and accuracy of the empirical conclusions previously drawn. The significant result of the variable representing the proportion of women on the board (fboard) when lagged, corroborates the proposition that an increase in the proportion of women serving on the board aids in the subsequent year's corporate financing activities. This finding complements the benchmark regression conclusions, suggesting that women's participation in significant decision-making and business supervision processes can also reduce the financing constraints faced by enterprises.

**Table 6. One-period-lagged Dependent Variables**

| VARIABLES | (1) kz | (2) kz | (3) kz |
|---|---|---|---|
| L.fceo | -0.320** | | |
|  | (-2.38) | | |
| L.fmanage | | -0.452*** | |
|  | | (-2.68) | |
| L.fboard | | | -0.579*** |
|  | | | (-2.73) |
| size | 0.215*** | 0.210*** | 0.216*** |
|  | (5.62) | (5.46) | (5.68) |
| roa | -12.570*** | -12.544*** | -12.568*** |
|  | (-31.47) | (-31.51) | (-31.54) |
| board | 0.375* | 0.341* | 0.345* |
|  | (1.89) | (1.73) | (1.75) |
| indep | 2.063*** | 1.991*** | 1.961*** |
|  | (3.03) | (2.94) | (2.89) |
| age | -0.242* | -0.227* | -0.229* |
|  | (-1.86) | (-1.74) | (-1.75) |
| Constant | -4.377*** | -4.147*** | -4.268*** |
|  | (-4.20) | (-3.96) | (-4.10) |
| | | | |
| Observations | 5,006 | 5,006 | 5,006 |
| R-squared | 0.331 | 0.331 | 0.331 |
| Industry FE | YES | YES | YES |
| Year FE | YES | YES | YES |

Robust t-statistics in parentheses
*** $p<0.01$, ** $p<0.05$, * $p<0.1$

Note: Table 6 performs a robustness check aims to account for the time-lagged effects of management changes and to alleviate potential endogeneity concerns in the empirical findings. Specifically, the newly significant result for fboard, with a lag, reinforces the hypothesis that the presence of women on the board contributes positively to the firm's financial activities in the following year. These findings substantiate the initial conclusions, reflecting a broader, time-extended impact on corporate finance.

## 5. Discussion

The study delves into the evolving role of female entrepreneurs and leaders in the corporate sphere, particularly in the context of China's dynamic market. It hypothesizes that female leadership in top management positions, including CEOs and board members, significantly influences a company's financial constraints. The methodology involves a comprehensive analysis of data from 938 entrepreneurial companies, focusing on various leadership roles including the presence of female CEOs (fceo), the proportion of women in top management (fmanage), and the presence of female board members (fboard). The study employs statistical regression models to analyze the data, along with Propensity Score Matching (PSM) and lagged dependent variables for eliminating selection bias and robustness checks. As such, the results indicate a clear and robust correlation between female leadership in CEO and senior management positions and reduced financial constraints within firms. Specifically, firms with female CEOs or a high proportion of women in top management roles exhibit significantly fewer financial constraints: the coefficients for fceo and fmanage are -0.318 and -0.326, respectively. However, the impact of female board members (fboard) on financial constraints was less definitive, as the estimated coefficient for fboard is -0.202, which, while negative, is not statistically significant. This suggests that the role of female board members in influencing financial constraints is less direct, potentially due to the nature of board functions which often involve decision-making on critical corporate matters or just having regular supervision, rather than direct involvement in daily management processes.

The study also uncovers industry-specific patterns, revealing stronger gender discrimination in management within high-tech industries. This discrimination potentially hinders the positive effects of female leadership on alleviating financing constraints in these sectors. The research contributes significantly to the discourse on gender dynamics in corporate finance, underscoring the need for greater female representation in leadership roles for improved corporate governance and financial health. This research also provides valuable insights into the role of gender dynamics in corporate financial management, particularly in the rapidly evolving Chinese economic landscape. It emphasizes the importance of promoting female leadership for not only advancing gender equity but also for enhancing the overall financial resilience of corporations. The findings are robust, validated through rigorous statistical methodologies, and contribute to a deeper understanding of the intersection between gender and corporate finance.